\documentclass[fleqn,usenatbib]{mnras}

\usepackage{newtxtext,newtxmath}
\usepackage[T1]{fontenc}

\DeclareRobustCommand{\VAN}[3]{#2}
\let\VANthebibliography\thebibliography
\def\thebibliography{\DeclareRobustCommand{\VAN}[3]{##3}\VANthebibliography}

\usepackage{graphicx}	% Including figure files
\usepackage{amsmath}	% Advanced maths commands

\title[Chemical evolution with mixing stochasticity]{Mixing stochasticity relinquishes evidence for magnetorotational hypernovae}

\author[A. Aggarwal et al.]{
Anmol Aggarwal,$^{1}$\thanks{E-mail: anmol.aggarwal.24@ucl.ac.uk}
Ralph Sch\"{o}nrich$^{1}$
\\
$^{1}$Mullard Space Science Laboratory, University College London, Holmbury St. Mary, Dorking, RH5 6NT, Surrey, UK\\
}

\date{Accepted XXX. Received YYY; in original form ZZZ}

\pubyear{\the\year{}}

\begin{document}
\label{firstpage}
\pagerange{\pageref{firstpage}--\pageref{lastpage}}
\maketitle
\defcitealias{yong2021r}{YK}
\defcitealias{aggarwal2026peculiar}{SA1}
\begin{abstract}

A recent work claimed by fitting the observed elemental abundance pattern of a halo star with the total yields from single super-/hypernova events that only a magnetorotational hypernova event (a very energetic supernova event that also produces r-process elements) could be the source of the observed abundances. Here, we show that the star's peculiar abundance pattern is better fitted within the framework of mixing stochasticity with yields from a normal core collapse supernova (ccSN; Energy\textsubscript{exp} \(\sim 10^{51}\) erg.) and a neutron star merger (NSM). The stochastic mixing model outperforms the hypernova fitting significantly, (r.m.s 0.24 vs 0.44) i.e., favours the composition from common events (ccSN + NSM) over the magnetorotational hypernova scenario. We also discuss the origin of the star and the possibility of enrichment of its birth cloud by both a ccSN and a NSM.
\end{abstract}


%%%%%%%%%%%%%%%%%%%%%%%%%%%%%%%%%%%%%%%%%%%%%%%%%%

%%%%%%%%%%%%%%%%% BODY OF PAPER %%%%%%%%%%%%%%%%%%

\section{Introduction}

While the abundances of today's Milky Way (MW) disc are remarkably homogeneous \citep{nieva2012present} with small expected peculiarities \citep{Freeman_and_bland_hawthorn_2002}, stars show increasing abundance scatter towards the lowest overall metallicities \citep{Francois2007}. It is well-known that stellar abundances track both the origin of stars and help us understand the physics of their host system. While modern abundances have been used to track, e.g., disc evolution, stellar radial migration, or also flows and accretion scenarios of the Galactic disc \citep{Goetz_and_Koeppen, Portimari_and_chiosi_1998, Schoenrich_and_Binney_2009, Bilitewski_and_schoenrich, Pezzulli_2016}, the early, very metal-poor stars found predominantly in the halo have triggered the creation of scenarios with low-number stochasticity and, in extreme cases, the strategy to fit single extremely metal-poor stars with single "progenitors" that polluted their parent cloud (e.g., \cite{Umeda_and_Nomoto_2003}). In a previous paper \citep{aggarwal2026peculiar} (hereafter \citetalias{aggarwal2026peculiar}), we catalogued the two main types of stochastic chemical evolution previously modelled: i) {\bf evolution stochasticity} \citep{colavitti2008}, which is concerned with changes in inflows and star formation history, and ii) {\bf Poisson stochasticity} \citep{karlsson2005, Koch2008, Venn, 10.1093/mnras/stab281}), which models the low number/Poisson noise of yield/supernova (SN) events in small systems. We showed that a third type of stochasticity matters: {\bf mixing stochasticity}, which accounts for the observed fact that SN explosions/remnants are highly inhomogeneous \citep{Hughes_2000, Fesen_2006, Larsson_2013} and can then (when the ISM is not completely mixed after enrichment) pollute neighbouring star-forming clouds with parts of SNe instead of the entire yield. In \citetalias{aggarwal2026peculiar} we showed that three stars (AS0039, HE 1327-2326, and J0931+0038) -- previously considered descendants of hypernovae (particularly energetic core collapse SN (ccSN) explosions with Energy\textsubscript{exp} $\gtrsim 10^{52}$ erg) -- are more easily explained by standard SN models with mixing stochasticity. We also concluded that our findings necessitate a re-analysis of all metal-poor/peculiar systems within this framework.

Here we examine a fourth peculiar star: SMSS J200322.54-114203.3 (hereafter SMSS 2003-1142), believed to be the descendant of a magnetorotational hypernova \citep{yong2021r}, hereafter \citetalias{yong2021r}. The star stands out against the previous objects, as it has high abundances of r-process elements in addition to the chemical peculiarity. It is a red giant, which by its orbital characteristics alone is clearly a member of the MW outer halo on a retrograde orbit, with iron abundance ([Fe/H]) $\sim -3.5$ dex (it lies amongst other outer halo objects in the $L_z$ vs. E diagram; see section \ref{arguments}). \citetalias{yong2021r} saw the peculiar abundances as a clear signature of a magnetorotational hypernova, as these are predicted to feature some r-process nucleosynthesis \citep{Symbalisty_1985, Winteler_2012}.

In this work, we, on the other hand, present an alternative case in which this star has not been enriched by a magnetorotational hypernova using prescriptions from \citetalias{aggarwal2026peculiar}, and there is no underlying unusual progenitor. Here we propose a scenario where a protostellar cloud enriched by a normal (Energy\textsubscript{exp} \(\sim 10^{51}\) erg) SN and by nucleosynthetic yields from a neutron star merger (NSM) produce an abundance pattern very close to the observations. We elucidate our argument using various statistical calculations and also r.m.s., which was used by \citetalias{yong2021r}.

Our paper is organised as follows: in the following section we discuss the origins of the star and list \citetalias{yong2021r}'s arguments; in section \ref{theory} we lay out the methods used for the SN + NSM fit; we then address the prognosis of NSM enrichment in section \ref{sec:nsm_ques}. After presenting the results in section \ref{res}, we tackle the arguments detailed in section \ref{arguments} and conclude in section \ref{conclusions}.

\section{SMSS 2003-1142}

\label{arguments}

SMSS 2003-1142, with Gaia identifier DR2/DR3 4190620966764303488, has galactic longitude $l \sim 30.3113$ deg and latitude $b \sim -21.1958$ deg with parallax $\varpi \sim 0.3998$ mas; rough estimates of its total energy and angular momentum suggest that this star is on a highly retrograde orbit, having a major fraction of the escape speed, and thus belongs to the outer halo: using a distance $\sim 2.5$ kpc and Solar motion of \cite{schoenrich2012}, we estimate a Local Standard of Rest centric velocity vector (U,V,W) $\sim$ (108.06122461153973 km/s, -469.6981987123922 km/s, -265.3412000074976 km/s) and a galactocentric velocity vector $(v_r, v_\phi, v_z)$ $\sim$ (-108.06122461153973 km/s -231.69819871239218 km/s -265.3412000074976 km/s). The circular speed has been taken as 238 km/s, and the Solar motion correction w.r.t. the Local Standard of Rest is (11.1 km/s, 12.24 km/s, 7.25 km/s). The overall assessment is consistent with \cite{Cordoni_gaia_2021}'s estimates. They used a significantly smaller distance and hence obtained a different velocity vector, while we use the Gaia DR3 parallax directly, and this distance is also compatible with \cite{bailer_jones_2021}.

Detailed abundances for this star have been measured by \citetalias{yong2021r} using the ultraviolet and visual echelle spectrograph (UVES) of the European Southern Observatory’s (ESO) Very Large Telescope (VLT). The star has an unknown age and a metallicity ([Fe/H]) of -3.5. The measured elemental abundances of SMSS 2003-1142 are listed in Table 1 of \citetalias{yong2021r}. We also report them in Table \ref{tab:obs} here.

In their original analysis, \citetalias{yong2021r} assert that this star has been enriched by a 25 \(M_\odot\) progenitor. While acknowledging that there are no magneto-rotational hypernova models or observations available in the present literature, they construct a model by combining nucleosynthesis yields from two different sources: i) \cite{nishimura2015r}'s B11$\beta$1.00 model, which is a post-processing nucleosynthesis calculation of a two-dimensional special-relativistic magnetohydrodynamic simulation of an Fe core from a rotating 25 \(M_\odot\) star at solar metallicity performed in \cite{takiwaki2009special}; ii) for the "envelope" (i.e., for elements with Z=1-32), a 25 \(M_\odot\) non-rotating, zero-metallicity, hypernova model with an explosion energy of $10^{52}$ erg, from a suite used in \cite{kobayashi2014origin} to explain the chemical abundances of some extremely metal-poor (EMP) stars.

\citetalias{yong2021r} use the root-mean-square (r.m.s.) statistic to determine the goodness of fit of their model to the data, so here we provide these in addition to further statistics.

Apart from model fitting, \citetalias{yong2021r} discuss several of these abundances and try to determine the properties exhibited by the progenitor of this star. They,
\begin{enumerate}

    \item deduce that a high [N/Fe] ratio relative to the [C/Fe] ratio indicates a fast-rotating progenitor.

    \item argue that high [Co/Fe] and [Zn/Fe] ratios of 0.36 and 0.72, respectively, can only be observed in the star if its progenitor was a high-energy SN (hypernova), even though abundances from C to Fe can be matched using normal SN yields.

    \item further rule out a SN enrichment by asserting that they can with a 15 \(M_\odot\) SN match the observed [Mg/Fe] ratio, but it would fall short on the observed [(Ca, Co, Zn)/Fe], and that the [Ni/Fe] ratio implies a magnetorotational hypernova, as its jet can eject the requisite Ni from near the black hole in their model.

    \item advocate the difficulty in explaining the low metallicity ([Fe/H]) of this C-normal star in a regime where multiple sources have enriched the star-forming ISM.

    \item state the impossibility of explaining Th abundance with a "spin star" model (i.e., a fast-rotating model) based on \cite{Choplin_2020}'s work.

    \item owing to the differences between the abundances of an r-process-enriched star (RAVE J183013.5-455510), for which a rotating massive star has been proposed as an r-process source, and SMSS 2003-1142's abundances (e.g., much lower C and higher Eu and Th), rule out this case.

    \item contend that the abundances of elements between C and Zn (except C, N, and Zn) reside within the distribution of abundance patterns observed in other EMP stars ([Fe/H] < -3) and this star is overall quite metal poor, and hence argue that a single enrichment event from a zero-metallicity ancestor is likely responsible for such an abundance pattern.

    \item stress that the observations for elements with Z < 31 are well explained by hypernovae from massive stars (M > 25 \(M_\odot\)), and for the r-process they need a magnetorotational SN; hence, a magnetorotational hypernova would be an apt progenitor.

    \item due to the low [Mn/Fe] ratio, state that type Ia SN could not have enriched it and there is no other evidence of it either.

    \item reject enrichment from type Ia SN, asymptotic giant branch stars, and electron-capture SN based on their galactic chemical evolution model in which the interstellar medium (ISM) reaches an [Fe/H] = -3.5 very quickly. (see section \ref{sec:nsm_ques} for an elaborate discussion on their model).

\end{enumerate}

While we agree with point (i) \citep{Meynet_2003}, i.e., favouring a fast-rotating model, the others can be easily mitigated in the realm of mixing stochasticity and some analytical arguments. As we show below, we discuss these arguments in section \ref{arg_adress} in detail. We do not contest the lack of SNIa in point (ix). However, for point (x), using metallicity as an indicator for stellar age appears unfounded to us. The star is clearly part of the outer halo, and this, together with the lack of an evident parent stream, points to a provenance from a very low-mass dwarf galaxy (and the kinematics exclude an in situ, or near-MW creation of the parent system). In turn, such systems can maintain very low metallicities on cosmic timescales – we revisit this in Section \ref{sec:nsm_ques}. Their point (viii) is negated directly by the fitting (presented in Fig. \ref{fig:full_set}) and the overall conclusions of this work.

\section{Underlying theory and methods}
\label{theory}

To fit the observed abundances with yield models, we adapt the approach from \citetalias{aggarwal2026peculiar}.

This mixing model implies the fitting equation from \citetalias{aggarwal2026peculiar}:

\begin{equation}\label{eq1}
\sum_{i,j=1}^{n,m} a_i x_{ij} = y_{j}+ \epsilon_j; a_i \in (0,1)
\end{equation}

where index \(i\) spans the different regions/shells the SN model has been divided into and \(j\) spans the different elements. $y$ is the observed abundance vector, and the $\epsilon_j$ is an error term, which budgets both model and observational uncertainty. $x_{i}$ is the abundance vector of each shell $i$. The model is solved for the coefficients \(a_i\), which are contributions of the shells.

In this work, we consider additional enrichment from an NSM, so we expand this equation to account for the additional yield vector $x_{nsm_{j}}$ from the NSM model and its contribution coefficient $a_{n+1}$:

\begin{equation}\label{eq2}
\sum_{i,j=1}^{n,m} a_i x_{ij}\  +\ a_{n+1} x_{nsm_{j}} = y_{j}+ \epsilon_j; a_i, a_{n+1} \in (0,1) 
\end{equation}

This model makes the simplifying assumption to sum over the NSM, i.e., to not account for eventual inhomogeneities in the NSM yields \citep{Wanajo_2014}.

For the NSM we use the yields from the SFHO\_13518 NSM model coupled with the remnant BH-torus model M3A8m1a5 of \cite{10.1093/mnras/stv009} obtained from the Garching Core-Collapse Supernova Archive \footnote{https://wwwmpa.mpa-garching.mpg.de/ccsnarchive/}. In accordance with the above expectation of a fast-rotating star, for the ccSN model, we use a 15 \(M_\odot\) normal SN model from \cite{limongi2018} which has a rotation speed of 150 km/s and an initial metallicity ([Fe/H]) of -3. As in \citetalias{aggarwal2026peculiar}, we dilute the ccSN + NSM contributions in a primordial composition gas cloud with 10 \(M_\odot\) of H, noting that the mass is of no importance other than the coefficients $a_i$ scaling proportionally.

We optimise the vector of coefficients ${\bf a}$ for the best possible fit to the data. We note one caveat here. Table \ref{tab:obs} lists two values of [Fe/H], one determined using neutral Fe lines and another using singly ionised Fe lines. \citetalias{yong2021r} mentions that to determine the [X/Fe] ratios for all the elements, they used [Fe/H] inferred from Fe I lines (denoted [Fe/H]$_I$) for neutral species and [Fe/H] inferred from Fe II lines ([Fe/H]$_{II}$) for singly ionised elements. As we use $N_x/N_H$ for $x_{ij}$, $x_{nsm_{j}}$, and $y_{j}$ in eqn. \ref{eq1}, we had to transform the values listed in Table \ref{tab:obs} to the [X/H] scale and determine the respective $N_x/N_H$ values for observations. I.e., following \citetalias{yong2021r}, we use two different bases for inferring the elemental abundances. The argument for this is to minimise systematic observation biases which could be common between the ionisation stages.

To get the final r.m.s. value consistent with \citetalias{yong2021r}, we use the difference between observations and the model for each [X/Fe] ratio. \citetalias{yong2021r} do not mention which solar abundances they refer to, but we found that \cite{asplund2009chemical} matches the dataset precisely and use the same throughout our analysis.

We use only the inner regions of the SN model (shown by the points in Fig. \ref{fig:coeff}) because with the lack of light element determinations, we have no observational constraints on contributions from the outermost shells.

\section{The NSM question}
\label{sec:nsm_ques}

The primary reason due to which \citetalias{yong2021r} reject enrichment from NSMs is because of their rate and delay time in their adopted galactic chemical evolution model. They emphasise that due to its high specific star formation rate, the metallicity of their MW model will so quickly surpass [Fe/H] > -3.5 dex that the minimum timescale needed for the first NSM will be undershot. While correct for their model, this ignores the fact that an object kpcs across cannot be mixed homogeneously on timescales of Myrs, and lower densities in outer regions will in turn imply slower enrichment timescales there. There is also a line of argument for neutron star binary systems having quick merger times \citep{matteucci_quick_nsm_2014, Cescutti_quick_nsm_2015}, but we do not rely on this argument.

Secondly, as pointed out by \cite{Schoenrich_and_weinberg} and also \cite{Fraser_and_schoenrich}, the question of the hot phase (receiving the bulk of stellar yields) vs. the cold star-forming phase of the ISM is central. Models that neglect this are expected to overestimate the early cold phase metallicity by about an order of magnitude. We also point out that the higher expected cold gas fraction of NSM yields, as argued by \cite{Schoenrich_and_weinberg}, would favour higher fractions of their yields being incorporated into star-forming clouds or, vice versa, allow for later NSM yields to still contribute significantly.

Most importantly, the above line of argument neglects the fact that \citetalias{yong2021r} classify in their own assessment SMSS 2003-1142 as a halo star and not an MW star. Its kinematics actually strongly argue against an early in situ formation, as the subsequent assembly of halo mass would force the star to lower actions/energy comparable with today's bulge/inner halo. Instead, the star has a galactocentric speed of about 368 km/s, placing it at much higher energy/speed than typical/inner halo stars, which implies a relatively late accretion to the MW. In addition, the star has a relatively large retrograde angular momentum, strengthening this case. In such small satellite/dwarf galaxies low iron abundances can persist on Gyr timescales. \citetalias{yong2021r} also acknowledge that inhomogeneous enrichment could, for dwarf galaxies and particularly ultra-faint dwarf galaxies, further favour metallicities below -3.5 dex for extended time periods.

In short, there appears to be no reason to a priori discard an NSM contribution for SMSS 2003-1142.

\section{results}
\label{res}

\begin{figure*}
    \centering
    \includegraphics[width=1\linewidth]{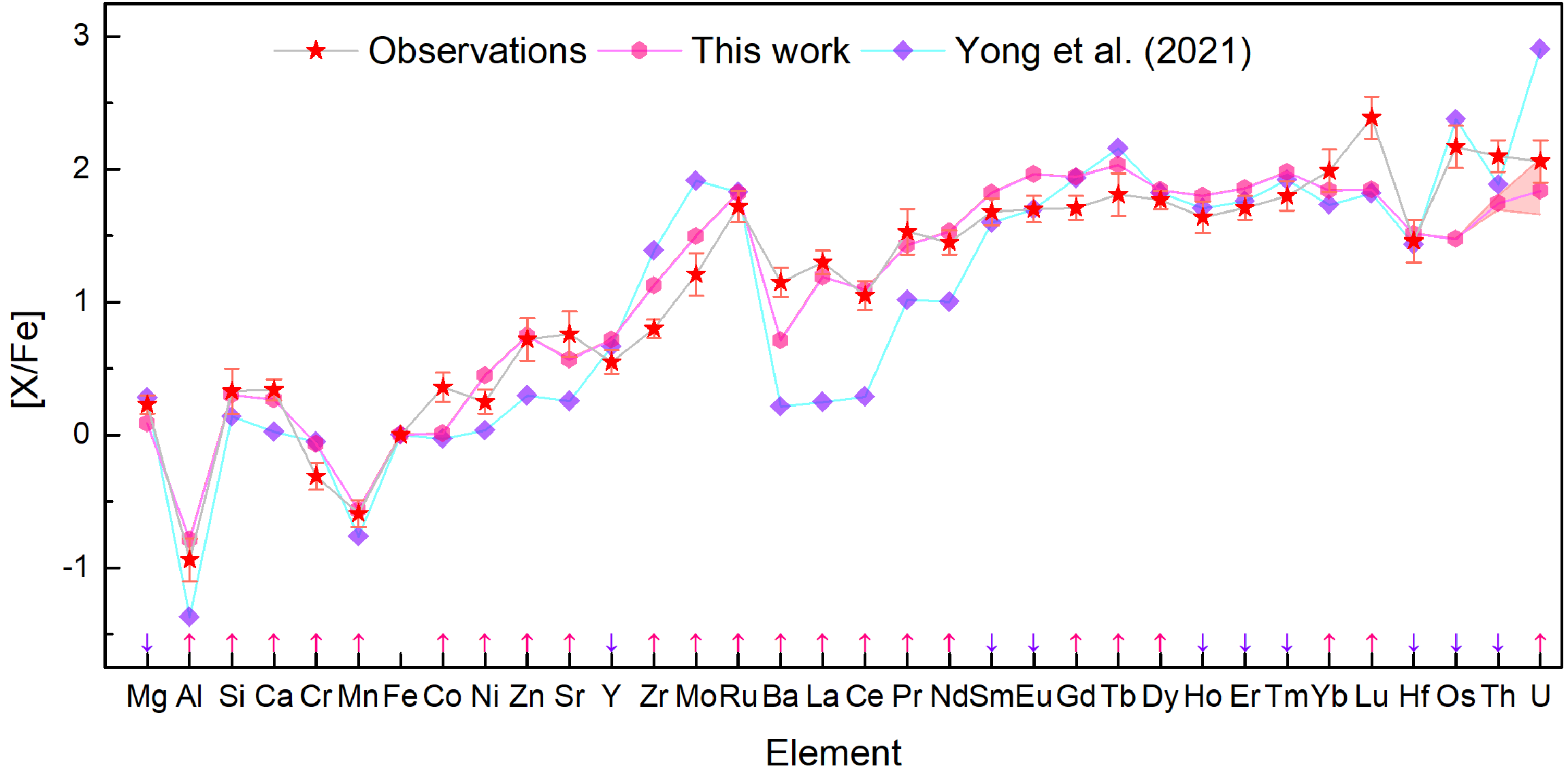}
    \caption{Comparison of \citetalias{yong2021r}'s magnetorotational hypernova fit and the SN + NSM fit to the full set of elements in \citetalias{yong2021r}. The arrows at the bottom encode which model (upward arrows for this work, downward for YK) is closer to the observations for each element.} (Fe is shown on the plot but has not been used during calculation of statistics)
    \label{fig:full_set}
\end{figure*}

\begin{figure}
    \centering
    \includegraphics[width=1\linewidth]{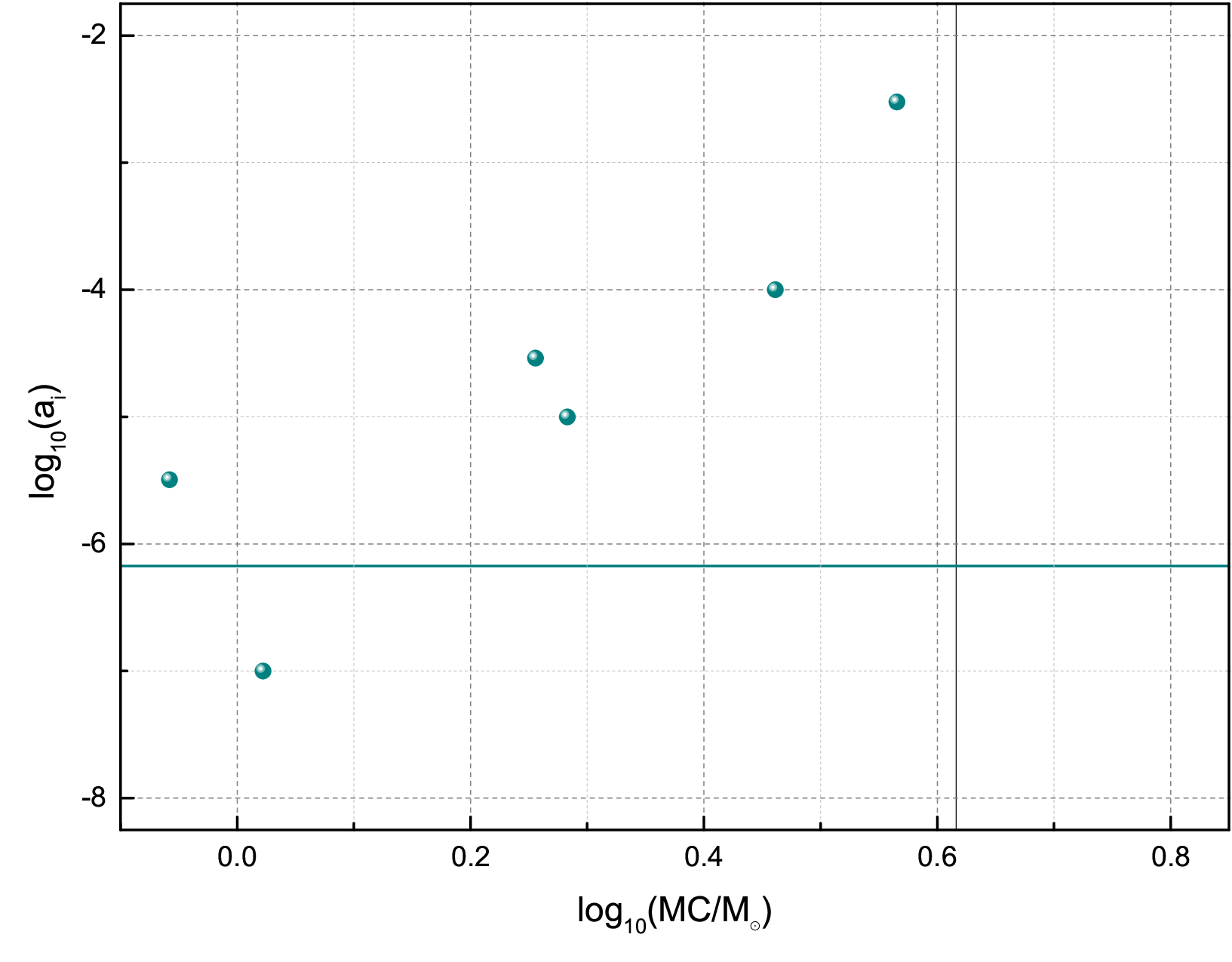}
    \caption{The mass coordinate (MC) of the 15 \(\textup{M}_\odot\) model star vs. the contribution coefficients (\(a_i\), on a log scale) resulting from fitting to the observations. The values of coefficients have been marked at the beginning of their respective regions. The solid vertical line marks the end of the last region. The solid horizontal line reports the NSM contribution. The underlying values have been given in Table \ref{tab:coeff_values}.}
    \label{fig:coeff}
\end{figure}

\citetalias{yong2021r} compare their magnetorotational hypernova fit to two different sets of observations, coining them "full" and "preferred" (without elements from Ba to Nd, inclusive) sets. They also exclude N, Na, Sc, Ti, and upper limits from both. For elements Ba to Nd, they ascribe the bad fits to uncertainties of their production in models which are in turn driven by model specifics. They excluded N and Na because their model is non-rotating and Sc and Ti, arguing their underproduction in one-dimensional SN models. The SN+NSM model, on the other hand, does not require the exclusion of any elements, and it has been fitted to the full set, enabling an r.m.s. much lower than the r.m.s. for both the full and preferred sets of \citetalias{yong2021r}.

They report an r.m.s of 0.44 and 0.34 for the full and preferred sets, respectively, in contrast to r.m.s obtained through our SN+NSM fit, which is 0.24 for the full set. Both the fits have been presented in Fig. \ref{fig:full_set}. For elements from Ba to Nd, which have a significant contribution from the s-process (Ba, La, and Ce being s-process-dominated and Pr and Nd being roughly equally contributed to by both the s- and r-processes), the SN+NSM model agrees with the observations much better than the hypernova model, which advocates for the need of some kind of s-process enrichment in the hypernova case and hence further weakens \citetalias{yong2021r}'s arguments. We also decayed the unstable isotopes in the model for 5 Gyrs: this introduces an additional uncertainty on Th ($\sim$ $\pm$ 0.05 dex) and U ($\sim$ $\pm$ 0.20 dex) model yields; everything else remains almost the same. The pink shaded area in Fig. \ref{fig:full_set} depicts this uncertainty bracket of 2.5 Gyrs to 7.5 Gyrs of decay. The observed abundances allow for two choices of argument: the decay of Th and U indicates a relatively young age if we trust the abundances; more likely, one can take the stance that the models are faulty, which is possible given the weakly constrained nuclear data, particularly at high nucleon numbers $A$. As discussed above, we tried to assess the kinematics of the star to obtain clues for the age – halo stars tend to be old, and limiting the accretion time of the parent system would establish a lower age limit. However, with the particularly high orbital energy and retrograde orbit, the star is placed in the outer Galactic halo, where late accretion is likely. Nevertheless, these elements are of lower importance: our conclusions remain unaffected if Th and U are taken out of the fitting. \citetalias{yong2021r} do not provide any information about how they handle the decay.

We note here that \citetalias{yong2021r}'s hypernova model values used in this analysis are only approximate, as they have been directly read from their Fig. 2(b). We calculated the r.m.s. using these values and obtained an r.m.s. of 0.44 dex, which \citetalias{yong2021r} reports; hence, any read-out errors are minimal. We discuss the statistics below and use a model error of 0.25 dex for all elements, which budgets these errors. We present in Fig. \ref{fig:CCSN_only} the contribution of the SN model to the overall fitting. It also depicts the s-process baseline.

The coefficients used for the SN and NSM models have been presented in Fig. \ref{fig:coeff}; the points represent the coefficients used for the SN shells and the mass coordinates of the shells used. The solid horizontal line corresponds to the coefficient used for the NSM model.

In addition to the r.m.s. statistic, we also calculate the $\chi^2$ and $\chi^2_{eff}$ for both the models using the following equations:

\begin{equation}
    \chi^2 = \sum_{i=1}^{n} \frac{\left( [X_i/Fe]_{\rm obs} - [X_i/Fe]_{\rm model} \right)^2}{(\sigma^2_{[X_i/Fe]_{\rm obs}} + \sigma^2_{[X_i/Fe]_{\rm model}})}
    \label{eq:chi_sq}
\end{equation}

\begin{equation}
    \chi^2_{eff} = \frac{\chi^2}{\nu} ; \nu = k - p
    \label{eq:sum_over_xi_fe}
\end{equation}

where \([X_i/Fe]_{\rm obs,model}\) and \(\sigma_{[X_i/Fe]_{\rm obs,model}}\) are the observed and theoretical elemental abundance ratios and standard deviations, respectively. $k$ and $p$ are the number of observations and the number of free parameters used in the models, respectively. For \citetalias{yong2021r} we take $p$ as 1 because of the fractional contribution from the neutron-rich ejecta to exactly match [Eu/Fe] in their fit. For the SN+NSM model we have $p$ = 7 (6 coefficients for the SN, 1 coefficient for the NSM). The dilution parameter is not taken as a free parameter here for either of the models because the fits are reported in [X/Fe] ratios and the $\chi^2$ is also being calculated using the same values. We assume \(\sigma_{[X_i/Fe]_{model}}\) = 0.25 for both \citetalias{yong2021r}'s and the SN+NSM models.

We present all the statistical measurements in Table \ref{tab:all_stats}. It is clear that the SN+NSM model outperforms the model in \citetalias{yong2021r} on all metrics. Interestingly, the SN+NSM model's $\chi^2_{eff}$ is close to 1, which indicates a very good fit, no overfitting, and no overestimated errors. We also calculate the p-value for both cases, i.e., testing the hypothesis that each model applies. Here, the respective p-values inform the probability of obtaining for an unbiased model a $\chi^2$ larger or equal to the measured value. 
The ratio of these p-values can be read as the likelihood ratio between the two models (modulo Bayesian prefactors), i.e. ascribing a much higher likelihood to the SN+NSM model.

We also note that in principle one could try to argue for larger model errors (e.g., pointing to uncertain nucleosynthesis chains), which has a direct impact on the $\chi^2$ and the ${\chi^2}_{eff}$. However, this is ruled out with the large number of elements available here. I.e., if we were to push up the model error, the p-value for the SN+NSM model would, e.g., rise to 0.9997 for a 0.5 dex model error, which rules out this error choice at the $0.001$ significance level.

Additionally, we calculate the error and Akaike information criterion-corrected Vuong's test (V\textsubscript{AIC \ \& \ {\rm error} \ {\rm corr.}}) to test if the added complexity is justified. The SN+NSM model comfortably crosses the 95\% threshold with a V\textsubscript{AIC \ \& \ error \ corr.} of -2.70, with the threshold being -1.96.

\begin{table}
    \centering
    \caption{Comparison of statistics for the Magnetorotational Hypernova model and SN+NSM model.}
    \begin{tabular}{ccccc}
    \hline
    \hline
        Fit/Statistic & r.m.s & $\chi^2$ & ${\chi^2}_{eff}$ & $p-value (\sigma = 0.25)$\\ \hline
        M.R. Hyp. (Full set) & 0.44 & 81.68 & 2.55 & $3.2 \times 10^{-6}$\\ \hline
        SN + NSM (Full set) & 0.24 & 26.54 & 1.02 & 0.43\\ \hline
        \multicolumn{2}{c}{V\textsubscript{AIC \ \& \ error \ corr.}} & \multicolumn{3}{c}{-2.70} \\ \hline
        
    \end{tabular}
    \label{tab:all_stats}
\end{table}

\section{Examination of arguments for a magnetorotational hypernova}
\label{arg_adress}

As signalled by the statistics and seen in Fig. \ref{fig:full_set}, for most elements on which \citetalias{yong2021r} base arguments for a magneto-rotational hypernova (below), the SN+NSM model performs better. The model presented here underpredicts [Co/Fe] by $\sim$ 0.35 dex vs. their $\sim$ 0.39 dex and [Zn/Fe] by $\sim$ 0.03 dex vs. their $\sim$ 0.42 dex; the high observed abundance of these two elements relative to standard/solar iron peak abundances points to high entropy \citep{umeda_and_nomoto_2002} and has been argued by \citetalias{yong2021r} to constitute evidence of hypernova enrichment (cf. argument (ii) in section \ref{arguments}). The SN+NSM model fits Ca, Mg, and Ni relatively well; for Ca, it fits signficantly better ($\Delta$[X/Fe] $\sim 0.07$ dex for SN+NSM vs. 0.32 dex for \citetalias{yong2021r}). It respectively under- and overpredicts Mg and Ni by $\sim$ 0.14 dex and $\sim$ 0.20 dex vs an overprediction of 0.05 dex for Mg and an underprediction of 0.22 dex for Ni in \citetalias{yong2021r}; we note here that the fitting of these elements can be further improved relatively easily by adjusting the SN shells/coefficients used in eqn \ref{eq1}, but we refrain from these minor adjustments to prevent overfitting. Nevertheless, the fits to Ca, Mg, and Ni along with fits to Co and Zn mitigate point (iii), which tried to rule out SN enrichment based on the fits to these elements.

For points iv to vi that aimed to discard multiple enrichment sources and (fast-) rotating models, if mixing stochasticity (\citetalias{aggarwal2026peculiar}) in enrichment is taken seriously, it will also imply that scenarios with multiple enrichment sources become more likely compared to single-enrichment scenarios. I.e., winds from several stars and parts of one or more SN of different kinds can enrich the same star-forming region and can reproduce the observed abundance pattern while keeping the metallicity low, provided they can also replicate the r-process elemental abundances along with normal nucleosynthesis products. However, better constraints on ISM mixing are expected to set lower limits on metallicity depending on the number of sources involved.
Further constraints on and tests of the mixing stochasticity can help in commenting on the possibility of the multiple enrichment scenario more robustly. We do not test these cases here, but it would be interesting to see what can be inferred by coupling mixing stochasticity to these.ece

For the elements fitted between C and Zn, SN+NSM fit has an r.m.s. of only $\sim$ 0.17 dex (i.e., comparable to the observational/model error) compared to theirs $\sim$ 0.30 dex. Hence the statement (vii) in the arguments of section \ref{arguments} , which advocated for enrichment from a single zero metallicity progenitor is dispelled, as evidently our model produces a better fit. For the point that other EMP stars show similar patterns and have a single progenitor, hence, SMSS 2003-1142 must have a single progenitor as well, bringing in r-process enrichment in addition to other nucleosynthesis products, we again point to the fact that mixing stochasticity allows for having more than one enrichment source for any EMP star. As in \citetalias{aggarwal2026peculiar}, this emphasises the need for quantification of this stochasticity and a reanalysis of all metal-poor stars in this new regime.

In addition, based on the analyses of just the elements between C and Zn, a second source only for r-process enrichment cannot be ruled out: i.e., an NSM. Even if one considers a single enrichment source for other EMP stars, it only implies a single SN progenitor for these elements and offers no constraints on r-process enrichment sources for SMSS 2003-1142 because an SN explosion does not contribute towards r-process enrichment. In any case, the fits presented in Fig. \ref{fig:full_set} along with the discussion in section \ref{sec:nsm_ques} already point towards two sources for this star.

These were the principal arguments drawn by \citetalias{yong2021r} to support their claim of a magnetorotational hypernova, which we mitigate easily.

\section{Conclusions}
\label{conclusions}

This work re-examines the origin of elements in a particular, r-process-enhanced star in the context of mixing stochasticity in chemical enrichment \citepalias{aggarwal2026peculiar}.

Our reanalysis changes the interpretation of this peculiar star. Previously thought to be clear evidence for enrichment by a magnetorotational hypernova \citepalias{yong2021r}, our new fit favours combined contributions from a normal SN and a neutron star merger (NSM), providing a significantly better match to the data ($\chi^2$ is 26.54 vs. 81.68 for the full element set, and r.m.s. is 0.24 vs. 0.44).

We note further that with this star, now the main $4$ objects deemed to be evidence for hypernova enrichment have all found a simpler explanation with a superior fit.

As discussed in \citetalias{aggarwal2026peculiar}, neither mixing efficiency within the SN nor mixing in the ISM is constrained by theoretical models. However, again, the derived contribution coefficients provide clues: while we did not impose any a priori constraints or merit functions, as in \citetalias{aggarwal2026peculiar}, the contributions show again a consistent and significant trend with radius (here outer shells increasingly contribute). This suggests significant but incomplete mixing in the ejecta. We stress again that this offers an interesting new route to explore and constrain the structure and dynamics of the ISM.

We note that the scenario suggested here (partial SN enrichment + NSM) is attractive, since NSMs are expected to contribute particularly well to the cold star-forming ISM directly (see e.g. \cite{Schoenrich_and_weinberg}). The kinematics and metallicity of this star point to a late accretion into the outer MW halo of a small system, and we stress that in a small parent galaxy the relative favouring of NSM yields should naturally give rise to r-process-enhanced stars such as this one, unless there is just one single star-formation spike.

We note that we tested one scenario here and that there could also be other scenarios at play, like enrichment from a more rapidly rotating progenitor or multiple enrichment sources, which can be tested under this same framework. So, while our statistics favour this SN and NSM combination over the previously considered hypernova case, we do not preclude other scenarios.

\section*{Acknowledgements}

We gratefully acknowledge the support from H.-Thomas Janka and Stephane Goriely for providing the NSM models and detailed discussions. The authors also thank Marco Limongi and Alessandro Chieffi for providing the explosive nucleosynthesis yields for their models used in this work. RS acknowledges the generous funding of a Royal Society University Research Fellowship. AA acknowledges funding from the University College London Research Excellence Scholarship.

%%%%%%%%%%%%%%%%%%%%%%%%%%%%%%%%%%%%%%%%%%%%%%%%%%
\section*{Data Availability}

The main statistics have been presented in the paper. We are happy to provide any additional information upon request.



\bibliographystyle{mnras}
\bibliography{example} % if your bibtex file is called example.bib

%\begin{thebibliography}{99}
%\end{thebibliography}

%%%%%%%%%%%%%%%%%%%%%%%%%%%%%%%%%%%%%%%%%%%%%%%%%%

%%%%%%%%%%%%%%%%% APPENDICES %%%%%%%%%%%%%%%%%%%%%

\appendix

\section{Measured abundances and additional fitting information}

\begin{figure}
    \centering
    \includegraphics[width=1\linewidth]{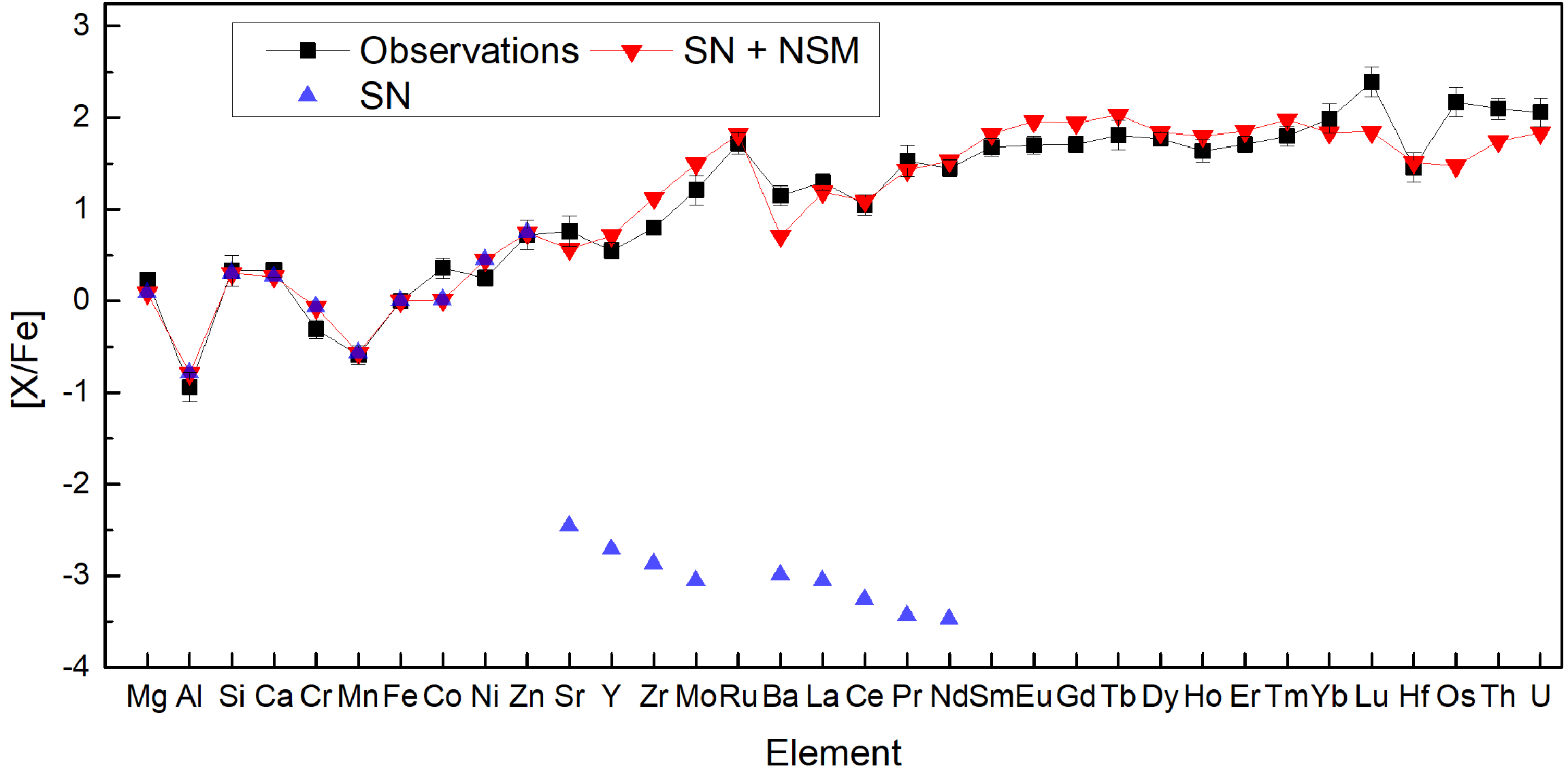}
    \caption{Comparison of the contribution by the SN model with the overall fitting. The blue upward triangles depict the SN contribution to the overall model (red downward triangles). As expected, most of the metals lighter than Zn are primarily contributed by the SN.}
    \label{fig:CCSN_only}
\end{figure}

\begin{table}
    \centering
    \caption{Elemental abundances of SMSS 2003-1142.}
    \label{tab:obs}
    {\fontsize{8.5}{2}\selectfont
    \begin{tabular}{|c|c|c|c|c|c|c|}
    \hline
    Species & Z  & log$\epsilon$(X) & $N_{\scriptscriptstyle \text{lines}}$ & s.e.m. & [X/Fe] & Total error \\ \hline
    C (CH)   & 6  & <4.93   & -        & -      & <0.07  & 0.3          \\ \hline
    N (NH)   & 7  & 5.33    & -        & -      & 1.07   & 0.3          \\ \hline
    Na I     & 11 & 2.80    & 2        & 0.09   & 0.13   & 0.11         \\ \hline
    Mg I     & 12 & 4.26    & 7        & 0.03   & 0.23   & 0.07         \\ \hline
    Al I     & 13 & 1.94    & 1        & -      & -0.94  & 0.16         \\ \hline
    Si I     & 14 & 4.27    & 1        & -      & 0.33   & 0.17         \\ \hline
    Ca I     & 20 & 3.11    & 7        & 0.02   & 0.34   & 0.08         \\ \hline
    Sc II    & 21 & 0.09    & 1        & -      & 0.37   & 0.19         \\ \hline
    Ti I     & 22 & 1.83    & 1        & -      & 0.45   & 0.17         \\ \hline
    Ti II    & 22 & 1.77    & 17       & 0.05   & 0.26   & 0.12         \\ \hline
    Cr I     & 24 & 1.76    & 5        & 0.07   & -0.31  & 0.1          \\ \hline
    Mn I     & 25 & 1.27    & 3        & 0.04   & -0.59  & 0.1          \\ \hline
    Fe I     & 26 & 3.93    & 91       & 0.02   & -3.57  & 0.11         \\ \hline
    Fe II    & 26 & 4.07    & 6        & 0.06   & -3.43  & 0.13         \\ \hline
    Co I     & 27 & 1.78    & 3        & 0.08   & 0.36   & 0.11         \\ \hline
    Ni I     & 28 & 2.90    & 3        & 0.07   & 0.25   & 0.09         \\ \hline
    Cu I     & 29 & <1.30   & 1        & -      & <0.68  & 0.16         \\ \hline
    Zn I     & 30 & 1.71    & 1        & -      & 0.72   & 0.16         \\ \hline
    Sr II    & 38 & 0.20    & 2        & 0.02   & 0.76   & 0.17         \\ \hline
    Y II     & 39 & -0.67   & 7        & 0.03   & 0.55   & 0.09         \\ \hline
    Zr II    & 40 & -0.05   & 12       & 0.02   & 0.80   & 0.07         \\ \hline
    Mo I     & 42 & -0.48   & 1        & -      & 1.21   & 0.16         \\ \hline
    Ru I     & 44 & -0.10   & 2        & 0.04   & 1.72   & 0.12         \\ \hline
    Rh II    & 45 & <-0.78  & 1        & -      & <1.74  & 0.16         \\ \hline
    Pd I     & 46 & <-0.72  & 1        & -      & <1.28  & 0.16         \\ \hline
    Ag I     & 47 & <-1.71  & 1        & -      & <0.92  & 0.16         \\ \hline
    Ba II    & 56 & -0.10   & 4        & 0.03   & 1.15   & 0.11         \\ \hline
    La II    & 57 & -1.03   & 7        & 0.02   & 1.30   & 0.09         \\ \hline
    Ce II    & 58 & -0.80   & 3        & 0.08   & 1.05   & 0.11         \\ \hline
    Pr II    & 59 & -1.18   & 1        & -      & 1.53   & 0.17         \\ \hline
    Nd II    & 60 & -0.56   & 8        & 0.03   & 1.45   & 0.09         \\ \hline
    Sm II    & 62 & -0.79   & 4        & 0.03   & 1.68   & 0.1          \\ \hline
    Eu II    & 63 & -1.21   & 4        & 0.02   & 1.70   & 0.1          \\ \hline
    Gd II    & 64 & -0.65   & 6        & 0.03   & 1.71   & 0.09         \\ \hline
    Tb II    & 65 & -1.32   & 1        & -      & 1.81   & 0.16         \\ \hline
    Dy II    & 66 & -0.56   & 14       & 0.02   & 1.77   & 0.07         \\ \hline
    Ho II    & 67 & -1.31   & 2        & 0.01   & 1.64   & 0.12         \\ \hline
    Er II    & 68 & -0.80   & 5        & 0.03   & 1.71   & 0.09         \\ \hline
    Tm II    & 69 & -1.53   & 3        & 0.09   & 1.80   & 0.11         \\ \hline
    Yb II    & 70 & -0.60   & 1        & -      & 1.99   & 0.16         \\ \hline
    Lu II    & 71 & -0.94   & 1        & -      & 2.39   & 0.16         \\ \hline
    Hf II    & 72 & -1.12   & 1        & -      & 1.46   & 0.16         \\ \hline
    Os I     & 76 & 0.00    & 1        & -      & 2.17   & 0.16         \\ \hline
    Pb I     & 82 & <-0.10  & 1        & -      & <1.72  & 0.16         \\ \hline
    Th II    & 90 & -1.31   & 2        & 0.09   & 2.10   & 0.12         \\ \hline
    U II     & 92 & -1.91   & 1        & -      & 2.06   & 0.16         \\ \hline
    \end{tabular}
    }
    \footnotesize{ $N_{\scriptscriptstyle \text{lines}}$, number of lines used to analyse the given species; s.e.m., standard error of the mean.}
\end{table}

\begin{table}
    \centering
    \caption{The mass coordinates (M.C.) of the SN shells and coefficients used for them and the NSM yields presented in Fig. \ref{fig:coeff}}
    \label{tab:coeff_values}
    
    \begin{tabular}{cccc}
    \hline
    \hline
         SN shell No./Source & Starting M.C. (\(M_\odot\)) & Ending M.C. (\(M_\odot\)) & Coeff. used\\ \hline
         1. & 0.87504 & 1.05239 & $3.2 \times 10^{-6}$\\
         2. & 1.05239 & 1.80176 & $1 \times 10^{-7}$\\
         3. & 1.80176 & 1.91828 & $2.9 \times 10^{-5}$\\
         4. & 1.91828 & 2.89178 & $1 \times 10^{-5}$ \\
         5. & 2.89178 & 3.67611 & $1 \times 10^{-4}$\\
         6. & 3.67611 & 4.13303 & $3 \times 10^{-3}$\\
         NSM & -- & -- & $6.7 \times 10^{-7}$\\
         \hline
    \end{tabular}
    \label{tab:placeholder}
\end{table}

%%%%%%%%%%%%%%%%%%%%%%%%%%%%%%%%%%%%%%%%%%%%%%%%%%

% Don't change these lines
\bsp	% typesetting comment
\label{lastpage}
\end{document}